\documentclass{iopjournal}

\usepackage{subcaption}
\usepackage{booktabs}
\usepackage[round]{natbib}
\usepackage{algorithm}
\usepackage{algpseudocode}
\usepackage{amsmath}
\usepackage{amsfonts}
\usepackage{todonotes}
\usepackage{siunitx}
\usepackage{enumitem}

\usepackage[acronym,shortcuts]{glossaries}
\newacronym{CGF}{CGF}{coverage-guided fuzzing}
\newacronym{DSE}{DSE}{deep stimulus encoder}
\newacronym{ML}{ML}{machine learning}
\newacronym{VO-KMVP}{VO-KMVP}{Violation-Output K-Multisection Violation Proportion}
\newacronym{VO-KMOC}{VO-KMOC}{Violation-Output K-Multisection Output Coverage}
\glsdisablehyper  

\usepackage{xcolor}

\begin{document}

\articletype{Paper} 

\title{Fuzzing the brain: Automated stress testing for the safety of ML-driven neurostimulation}

\author{Mara Downing$^{1,*}$\orcid{0009-0006-8431-6695}, 
    Matthew Peng$^1$\orcid{0009-0008-2945-2170}, 
    Jacob Granley$^1$\orcid{0000-0002-9024-2454}, 
    Michael Beyeler$^{1,2}$\orcid{0000-0001-5233-844X}, and 
    Tevfik Bultan$^{1}$\orcid{0000-0003-2993-1215}
}

\affil{$^1$Department of Computer Science, University of California, Santa Barbara, CA, USA} \\
\affil{$^2$Department of Psychological \& Brain Sciences, University of California, Santa Barbara, CA, USA}


\affil{$^*$Author to whom any correspondence should be addressed.}

\email{maradowning@cs.ucsb.edu}


\begin{abstract}
\emph{Objective:} Machine learning (ML) models are increasingly used to generate electrical stimulation patterns in neuroprosthetic devices such as visual prostheses. While these models promise precise and personalized control, they also introduce new safety risks when model outputs are delivered directly to neural tissue. We propose a systematic, quantitative approach to detect and characterize unsafe stimulation patterns in ML-driven neurostimulation systems.
\emph{Approach:} We adapt an automated software testing technique known as \emph{coverage-guided fuzzing} to the domain of neural stimulation. Here, fuzzing performs stress testing by perturbing model inputs and tracking whether resulting stimulation violates biophysical limits on charge density, instantaneous current, or electrode co-activation. The framework treats encoders as black boxes and steers exploration with coverage metrics that quantify how broadly test cases span the space of possible outputs and violation types.
\emph{Main results:} Applied to deep stimulus encoders for the retina and cortex, the method systematically reveals diverse stimulation regimes that exceed established safety limits. Two violation-output coverage metrics identify the highest number and diversity of unsafe outputs, enabling interpretable comparisons across architectures and training strategies.
\emph{Significance:} Violation-focused fuzzing reframes safety assessment as an empirical, reproducible process. By transforming safety from a training heuristic into a measurable property of the deployed model, it establishes a foundation for evidence-based benchmarking, regulatory readiness, and ethical assurance in next-generation neural interfaces.
\end{abstract}

\section{Introduction}
\label{sec:introduction}

\Ac{ML} is rapidly transforming neuroengineering by enabling adaptive encoding and decoding of neural activity in systems that restore or augment human function. 
In visual prostheses~\citep{fernandez2018development,ayton_update_2020}, deep neural networks have been proposed to translate camera images into electrical stimulation patterns delivered to the retina or cortex~\citep{granley2022hybrid,de_ruyter_van_steveninck_end--end_2022,granley_human---loop_2023,moure_deep_2025}. 
These networks implement the inverse of a forward model~\citep{chen_simulating_2009,beyeler2019model,granley_computational_2021,granley_adapting_2022,van2024towards}, which maps electrical stimulation to predicted neural or perceptual responses.
Inverting this mapping yields a \emph{stimulus encoder} that transforms a desired percept (or its visual proxy) into the per-electrode stimulation patterns expected to elicit it.
%
Learned stimulus encoders are now being explored in early-stage clinical evaluations~\citep{moure_deep_2025} and are central to designs for next-generation prosthetic vision systems~\citep{beyeler_towards_2022,grani_toward_2022,grani2025neural}. 
Because they would prescribe electrical stimuli in real time, their outputs must adhere to established limits on charge density, instantaneous current, and active electrode count~\citep{park2018methodologic}. Ensuring adherence to these constraints is therefore a prerequisite for clinical translation.

However, the safety of stimulus encoder systems remains critically understudied. Typically, firmware or hardware safeguards built into the clinical system are relied upon to clip, rescale, or drop unsafe stimuli~\citep{second_sight_argus_2013}. While this prevents immediate harm to the user, it obscures whether the underlying encoder performed safely, and prevents iterative refinement to improve model safety while maintaining performance. The limited existing work in this area has aimed to reduce violations by penalizing unsafe stimuli during training \citep{kuccukouglu2025end}, but there remains a lack of tools that clinicians and researchers can use to systematically validate that model-generated stimuli adhere to safety constraints.

Here we introduce an automated stress-testing framework for evaluating the safety of \ac{ML}-driven neurostimulation. The approach adapts \emph{\ac{CGF}}~\citep{chen2018systematic} to probe encoders for unsafe output regimes. In this setting, fuzzing perturbs input images while monitoring whether the resulting stimulation exceeds predefined limits on charge density, instantaneous current, or the number of active electrodes. Exploration is directed by coverage signals that quantify how broadly the perturbations probe the encoder's output space and its proximity to safety boundaries, enabling the systematic discovery of rare but clinically meaningful failure modes.

To guide this process, we introduce two complementary output-space coverage metrics. The first, \emph{\ac{VO-KMVP}}, prioritizes tests that push stimulation parameters toward their physiological limits, revealing the inputs that provoke the most severe violations. The second, \emph{\ac{VO-KMOC}}, measures how broadly the test exercises the range of possible stimulation patterns across electrodes, emphasizing the diversity of violation types and spatial distributions. Together, these metrics characterize both the frequency and the breadth of unsafe behaviors.

We demonstrate this framework on state-of-the-art stimulus encoders for retinal and cortical prostheses~\citep{granley_human---loop_2023,van2024towards}, which were trained to optimize perceptual fidelity but not explicitly constrained for safety.
The stress test uncovers over-limit stimulation patterns that conventional testing does not effectively discover, offering quantitative insights that can guide model selection, retraining, and firmware policy. 
We then demonstrate how coverage-guided fuzzing can be used in conjunction with performance metrics to evaluate the safety improvements from different regularization strategies~\citep{kuccukouglu2025end}, showing its value as a tool to inform model selection and refinement.

Although our experiments focus on artificial vision, the same principles apply to any neural interface where an \ac{ML} model prescribes electrical stimulation under biophysical constraints, including next-generation deep brain, spinal, and vagus nerve stimulators~\citep{shenoy_combining_2014,rao_towards_2019,drakopoulos2023neural,okorokova2018biomimetic}.
By framing safety evaluation as output-level verification and validation, coverage-guided stress testing offers a generalizable foundation for developing safer and more trustworthy \ac{ML}-based neurotechnologies.

\begin{figure}[!tb]
    \centering
    \includegraphics[width=\linewidth]{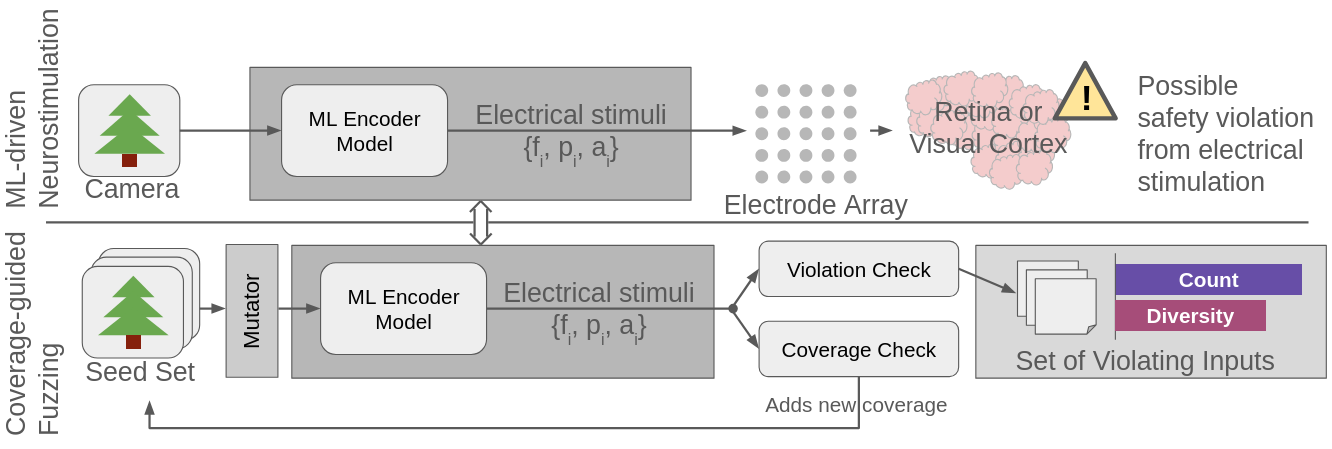}
    \caption{Overview of our framework for discovering safety violations in ML-driven neurostimulation. \emph{Top:} Visual prostheses use deep nets to convert camera input into electrical stimuli applied to the brain. Outputs must satisfy neurobiological constraints; violations may occur even under normal input. \emph{Bottom:} Our coverage-guided fuzzer mutates inputs to explore model behavior, using coverage and violation checks to uncover diverse unsafe outputs. The resulting violations enable quantitative safety evaluation and model comparison.}
    \label{fig:enter-label}
\end{figure}



\section{Methods}
\label{sec:methods}

Our goal is to systematically test whether a trained stimulus encoder ever produces stimulation parameters that exceed established biophysical limits. 
To do so, we adapt a software testing strategy called \acf{CGF}~\citep{chen2018systematic} to the domain of neurostimulation. 
In conventional software testing, fuzzing automatically perturbs program inputs to uncover rare failure modes; here, it perturbs sensory inputs (e.g., images) to expose conditions under which an \ac{ML} encoder produces unsafe stimulation. 
This allows the model to be evaluated in a \emph{black-box} fashion (i.e., no internal weights or gradients are needed) and complements the usual forward simulations or loss-based analyses used in model development.

We focus on regression models that map sensory input $\mathbf{x}\in\mathbb{R}^{d}$ (where $d$ indicates the number of dimensions in the input) to a vector of stimulation parameters $\mathbf{y}=M(\mathbf{x})$. 
In a neurostimulation system with $|\mathcal{I}|$ electrodes, the model output can be expressed as
$\mathbf{y}=\{f_i,p_i,a_i\}_{i=1}^{|\mathcal{I}|}$, where $f_i$ denotes pulse frequency, $p_i$ the pulse duration, and $a_i$ the amplitude of a biphasic square-wave pulse train delivered by electrode $i$. 
Safety is characterized by a set of inequality constraints $\{V_{k}(\mathbf{y}) \le 0\}_{k=1}^{K}$, each corresponding to a physiological limit (e.g., maximum charge density, instantaneous current, or co-activation area). 
An input $\mathbf{x}$ constitutes a \emph{violation input} if its output violates at least one constraint.

Formally, we aim to discover a large and diverse set of inputs
\begin{equation}
    \mathcal{V}=\{\mathbf{x}\mid \exists k: V_{k}(M(\mathbf{x}))>0\},    
\end{equation}
subject to the domain-specific constraints $V_k(\cdot) \le 0$. 
Each $V_k$ may apply globally (e.g., total current across all electrodes) or locally (e.g., per-electrode charge density), as detailed in Section~\ref{sec:violations}.

Because a single violation type can dominate the search, we guide exploration using coverage metrics that favor both the discovery of new violations and the diversification of test cases (Section~\ref{sec:coveragemetrics}). 
This balance ensures that the framework not only maximizes the number of unsafe cases found but also explores the search space, providing actionable insight for model redesign or retraining.

\subsection{Safety Constraints for Electrode-Based Neurostimulation}
\label{sec:violations}

Electrical stimulation delivered through implanted electrodes must obey strict biophysical limits to prevent tissue damage and patient discomfort. 
Typical devices control three parameters per electrode (i.e., the pulse frequency $f_i$, duration $p_i$, and amplitude $a_i$ of charge-balanced biphasic pulse trains) and safe operation requires that each combination remain within established physiological and device-specific bounds. 
Our framework treats these limits as formal constraints on the outputs of a model and identifies any violation of them as a potential safety risk.

We categorize violations into two broad types. 
\emph{Aggregate} violations occur when a property of the stimulation pattern as a whole exceeds a system-wide limit, such as total instantaneous current across all electrodes. 
\emph{Electrode-wise} violations occur when a single channel violates a local constraint, such as charge density or pulse timing. 
Formally, we express these as inequalities over the model’s output vector $\mathbf{y}$: a configuration is safe when all constraints $V_k(\mathbf{y}) \le 0$ are satisfied and unsafe when at least one $V_k(\mathbf{y})>0$.

Within this schema, we define four clinically motivated safety constraints representative of real retinal and cortical prostheses:
\begin{itemize}[topsep=0pt,itemsep=-1ex,partopsep=0pt,parsep=1ex]
    \item \emph{Physically impossible stimulus:} Each biphasic pulse must fit within its temporal period defined by its frequency $f_i$ (\SI{}{\hertz}). When the pulse duration $p_i$ (\SI{}{\milli\second}) becomes too long to complete a full cycle, the pulse is physically infeasible:
    \begin{equation}
        V_{PI} = 2p_i - \frac{1000}{f_i}.
        \label{eq:V_PI}
    \end{equation}
    
    \item \emph{Charge density limit:} To avoid electrochemical damage at the electrode–tissue interface, the delivered charge per electrode must remain below a device-specific limit. For epiretinal implants such as the Argus II, this limit is specified in the surgical manual~\citep{second_sight_argus_2013} as a \emph{per-electrode} maximum charge (derived from the FDA charge-density threshold and the device's electrode geometry). Accordingly, we treat the product of pulse duration $p_i$ and amplitude $a_i$ as a per-electrode charge quantity that must not exceed the published limit $\epsilon_1$:
    \begin{equation}
        V_{CD} = p_i a_i - \epsilon_1,
        \label{eq:V_CD}
    \end{equation}
    where a positive value indicates a violation.
    
   \item \emph{Instantaneous current limit:}
    The total instantaneous current across all electrodes $\mathcal{I}$ must stay below a device-level ceiling $\epsilon_2$ (\SI{}{\micro\ampere}), ensuring hardware stability and avoiding unintended current spread:
    \begin{equation}
        V_{IC} = \sum_{i=1}^{|\mathcal{I}|} a_i - \epsilon_2.
        \label{eq:V_IC}
    \end{equation}
    
    \item \emph{Active electrode limit:}
    The number of simultaneously active electrodes must remain below $\epsilon_3$ to minimize crosstalk and power consumption (here $[\cdot]$ denotes the Iverson bracket, equal to~1 if the condition inside is true and 0 otherwise):
    \begin{equation}
        V_{AE} = \sum_{i=1}^{|\mathcal{I}|} [a_i > 0] - \epsilon_3.
        \label{eq:V_AE}
    \end{equation}
\end{itemize}

In all cases here, a positive value ($V>0$) denotes a violation. The specific values used for $\epsilon_1$, $\epsilon_2$, and $\epsilon_3$, were derived separately for retinal and cortical prostheses using published literature and FDA specifications in conjunction with consultation with clinical experts \citep{fernandez_visual_2021, fernandez2018development, chen2020shape, fernandez2016cortivis, second_sight_argus_2013, fda_humanitarian_exemption}.

Although these expressions are taken from visual prosthesis designs, the same formulation applies to any electrode-based neurotechnology (e.g., cochlear, spinal, deep brain, or vagus nerve stimulators) where continuous control of amplitude, frequency, and pulse width must remain within safe biophysical limits~\citep{mccreery_charge_1990,grill_stimulus_1995,shannon_model_1992,cameron_safety_2004}. 
Defining safety directly in terms of model outputs allows our framework to evaluate encoder models in a black-box manner, independent of input type or behavioral context.

\subsection{Coverage‑Guided Fuzzing}
\label{sec:fuzzing}

In traditional software testing, fuzzing repeatedly perturbs program inputs to uncover rare failures such as crashes. 
Here, \ac{CGF} serves as an \emph{automated stress test}: the algorithm perturbs sensory inputs (e.g., camera images), observes the resulting stimulation patterns, and records any cases that violate the safety constraints defined in Section~\ref{sec:violations}. 
This process requires no access to model internals, making it well-suited for validation of proprietary or closed-source encoders.

A coverage function $\mathrm{Cov}(T)\!\in\![0,1]$ quantifies how much of the model’s behavioral space has been explored by a set of test inputs $T$. 
Coverage can be based on different signals (e.g., internal activations, output statistics, violation distributions; detailed in Section~\ref{sec:coveragemetrics}), but the goal is the same: higher coverage means a broader sampling of possible model behavior. 
The fuzzer begins with a seed set $S$ of initial test images which are iteratively mutated to produce new tests. 
A new test input $\mathbf{x}'$ is added to $S$ only if it increases coverage, that is, when $\mathrm{Cov}(S \cup \{\mathbf{x}'\}) > \mathrm{Cov}(S)$. 
In this way, the algorithm automatically steers exploration toward novel and potentially unsafe regions of model behavior.

\subsubsection{Fuzzing Strategy.}
The high-level procedure is summarized in Algorithm~\ref{alg:fuzzing}.
Before fuzzing, an optional preprocessing step estimates the expected range of input or output values, if required by the coverage metric. 
The algorithm then enters an iterative loop: it selects seed images, applies random perturbations (``mutations''), evaluates the resulting model outputs, and updates both the coverage and the list of discovered violations as necessary.

\begin{algorithm}[h!]
{\small
\caption{\textsc{Fuzz($S$, $P$)} \\ 
\Comment{Performs fuzzing on the model to detect safety violations.}\\ 
\Comment{Calls function \textsc{PreProcess}() which computes expected ranges for nodes or output values if required by the coverage metric, function \textsc{Cov}() which returns the model coverage as a value between 0 and 1, and function \textsc{TestMutants}() which is described in Algorithm~\ref{alg:mutating}.}}\label{alg:fuzzing}
\begin{flushleft}
  \textbf{Input:} $S$: seed set and $P$: optional set of input data for pre-processing.\\
  \textbf{Output:} $\mathcal{V}$: set of violating inputs.
\end{flushleft}
\begin{algorithmic}[1]
\State $\textsc{PreProcess}(P)$ \Comment{Computed values are stored globally}
\State $C \leftarrow \textsc{Cov}(S)$
\State $\mathcal{V} \leftarrow \emptyset$
\State $\textit{numberOfTests} \leftarrow |S|$
\While{$\textit{numberOfTests} < \textit{testLimit}$}
    \State $(S,\mathcal{V},C) \leftarrow\textsc{TestMutants}(S,\mathcal{V},C, m)$
\Comment{Generates and tests $m$ mutants, Algorithm~\ref{alg:mutating}}
    \State $\textit{numberOfTests} \leftarrow \textit{numberOfTests} + m$ 
\EndWhile
\State \Return $\mathcal{V}$
\end{algorithmic}
}
\end{algorithm}

\subsubsection{Mutation Strategy.}
Each new test input is generated by applying a random image-level transformation to a seed example, following image transformations from and procedures similar to DeepHunter~\citep{xie2019deephunter}. 
Transformations include translation, rotation, scaling, shearing, brightness or contrast adjustment, blurring, additive noise, and pixel-level perturbation. 
At each iteration, the algorithm:
\begin{enumerate}[topsep=0pt,itemsep=-1ex,partopsep=0pt,parsep=1ex]
    \item selects a seed $\mathbf{x}\!\in\!S$ for mutation, weighted by how often it has previously led to new violations (Eq.~\ref{eq:weight}),
    \item applies a random transformation to create $\mathbf{x}'$ and obtains $\mathbf{y}'\!=\!M(\mathbf{x}')$,
    \item checks whether $\mathbf{y}'$ violates any safety constraint $V_k(\mathbf{y}')$ and, if so, records $\mathbf{x}'$ in $\mathcal{V}$,
    \item evaluates whether $\mathbf{x}'$ increases coverage; if yes, it is added to the seed set $S$ for further exploration.
\end{enumerate}

This procedure, summarized in Algorithm~\ref{alg:mutating}, repeats for a fixed number of mutations per seed ($m\!=\!10$ in our experiments), progressively building a diverse collection of unsafe examples while exploring the available search space. The following equation controls how seeds are weighted for selection at each iteration of the algorithm:
\begin{equation}
P(s) = 
\begin{cases} 
1 - g(s)/\gamma, & \text{if } g(s) > 1 - p_{\mathrm{min}}\gamma,\\
p_{\mathrm{min}}, & \text{otherwise,}
\end{cases}
\label{eq:weight}
\end{equation}
where $P(s)$ is the probability of selecting seed $s$, $g(s)$ counts its prior selections, $\gamma$ scales sampling frequency, and $p_{\mathrm{min}}$ prevents any seed from being permanently ignored. This equation is adapted from Deephunter~\citep{xie2019deephunter}.

\begin{algorithm}
{\small
\caption{\textsc{TestMutants($S$, $\mathcal{V}$, $C$, $m$)} \\ 
\Comment{Generates $m$ mutant images from seed $s$, checks violations, and adds each one to the seed set if coverage is increased.}\\
\Comment{Calls function \textsc{Choose}() which chooses a seed as described in Eq. (\ref{eq:weight}) function \textsc{Mutate}() which chooses a mutation at random, applies it, and returns the new image, function \textsc{Cov}() which returns the model coverage as a proportion between 0 and 1, and function \textsc{Violates}() which returns a boolean indicating whether or not a test produces a violation.}}\label{alg:mutating}
\begin{flushleft}
  \textbf{Input:} $S$: seed set and $\mathcal{V}$: set of violating inputs found. $C$: current proportion of coverage using existing tests in $S$. $m$: number of mutants to generate.\\
  \textbf{Output:} $S$: new seed set (may be unchanged), $\mathcal{V}$: new list of violations (may be unchanged), and $C$: current proportion of coverage using existing tests in $S$.
\end{flushleft}
\begin{algorithmic}[1]
\State $s \leftarrow \textsc{Choose}(S)$ \Comment{$s$ is chosen from $S$}
\For{1 to $m$}
    \State $\mathbf{x}' \leftarrow\textsc{Mutate}(s)$
    \If{$\textsc{Cov}(S \cup \mathbf{x}') > C$}
        \State $S \leftarrow S \cup \mathbf{x}'$
        \State $C \leftarrow \textsc{Cov}(S)$
    \EndIf
    \If{$\textsc{Violates}(\mathbf{x}')$}
        \State $\mathcal{V} \leftarrow \mathcal{V} \cup \mathbf{x}'$
    \EndIf
\EndFor
\State \Return $(S,\mathcal{V},C)$
\end{algorithmic}
}
\end{algorithm}

\subsection{Coverage Metrics}
\label{sec:coveragemetrics}


Effective coverage-guided fuzzing requires a feedback signal that reflects how much of a model's behavior has been explored. 
This feedback is called \emph{coverage}. 
In conventional software testing, coverage often counts which lines of code were executed by a test set. 
For neural networks, prior work has used neuron activations as a stand-in for lines of code~\citep{pei2017deepxplore,ma2018deepgauge}, but such internal signals often fail to correlate with meaningful conclusions about model \emph{outputs}~\citep{yang2022revisiting,li2019structural,dong2020empirical,huang2024neuron}. 

In the context of neural stimulation, a good coverage metric should encourage the fuzzer to generate new tests that reveal \emph{distinct and physiologically relevant} stimulation patterns—those that either approach the boundaries of safe operation or differ meaningfully in their output configuration. 
Without such a signal, the fuzzer would produce redundant test cases or fail to uncover rare unsafe conditions.

To systematically investigate which coverage strategies best uncover safety violations, we evaluate eleven metrics grouped into three conceptual families:
\begin{itemize}[topsep=0pt,itemsep=-1ex,partopsep=0pt,parsep=1ex]
    \item \emph{Basic strategies:} simple heuristics that use no coverage signal. They serve as baselines, measuring the effect of naive approaches to utilizing mutations.
    \item \emph{Neuron coverage metrics:} white-box approaches that track how many internal neurons are activated by a test. These methods, adapted from software fuzzing for image classifiers, provide a historical reference but are not aligned with safety outcomes.
    \item \emph{Violation-focused metrics:} new black-box metrics we introduce that operate directly on model inputs and outputs, guiding the search toward diverse and physiologically meaningful safety violations. 
\end{itemize}

Table~\ref{tab:coverage_metrics} summarizes all eleven metrics, which are described in detail below.
The upper categories list the basic and neuron-based metrics used for comparison, while the lower category presents our six proposed violation-focused metrics.

\begin{table}[ht!]
{
  \caption{Fuzzing Coverage Metrics Used in This Paper}
  \label{tab:coverage_metrics}
  \centering
  \begin{tabular}{p{2.5cm}p{0.7\linewidth}}
    \hline
    \multicolumn{2}{l}{\textbf{Basic Strategies:}} \\
    \textbf{B-N} & Mutates user-provided seeds but does not add new tests to the seed set. \\
    \textbf{B-A} & Mutates and adds all new tests to the seed set. \\
    \textbf{B-FR} & Uses fully random images without mutation or a seed set. \\
    \textbf{B-Local} & Perturbs the seed with the most violations locally to generate many similar tests. \\
    \hline
    \multicolumn{2}{l}{\textbf{Neuron Coverage Metrics (White-Box):}} \\
    \textbf{N-NC} & Activates neurons exceeding a fixed threshold~\citep{pei2017deepxplore}. \\
    \textbf{N-KMNC} & Partitions each neuron's output into $K$ bins and tracks which are activated~\citep{ma2018deepgauge}. \\
    \textbf{N-NBC} & Tracks activations above or below neuron-specific bounds from training data~\citep{ma2018deepgauge}. \\
    \textbf{N-SNAC} & Tracks activations that exceed the maximum seen in training data~\citep{ma2018deepgauge}. \\
    \textbf{N-TKNC} & Tracks top-$K$ most activated neurons per layer~\citep{ma2018deepgauge}. \\
    \hline
    \multicolumn{2}{l}{\textbf{Novel Violation-Focused Coverage Metrics (Black-Box):}} \\
    \textbf{\ac{VO-KMVP}} & Bins the proportion of violation severity (including no violation) for each constraint. \\
    \textbf{\ac{VO-KMOC}} & Bins each output dimension across the test set. \\
    \textbf{\ac{VO-KMVP}-V} & Like \ac{VO-KMVP}, but only considers proportions which indicate a violation. \\
    \textbf{VO-VCC} & Tracks which constraints have been violated at least once. \\
    \textbf{I-KMIC} & Bins each pixel's value range across the input. \\
    \textbf{I-Div-Approx} & Bins feature space from an autoencoder to approximate test diversity. \\
    \hline
  \end{tabular}}
\end{table}

\subsubsection{Design Rationale}
\label{sec:metrics-rationale}

Our goal is not to invent arbitrary metrics, but to span the most plausible design space for coverage in this domain. 
We systematically explored metrics that operate in three spaces relevant to an encoder model:
\begin{itemize}[topsep=0pt,itemsep=-1ex,partopsep=0pt,parsep=1ex]
    \item \emph{Input space:} encouraging diverse sensory inputs,
    \item \emph{Feature space:} encouraging diversity in latent features,
    \item \emph{Output and violation space:} encouraging exploration of stimulation patterns and safety limits.
\end{itemize}
This principled organization ensures that our proposed metrics cover every meaningful axis along which coverage-guided exploration might improve safety testing.

\subsubsection{Violation-Focused Metrics (Our Approach)}
\label{sec:metrics-ours}

Among the new metrics, two perform consistently best and form the core of our framework: 
\emph{\acf{VO-KMVP}} and 
\emph{\acf{VO-KMOC}}.
Both metrics quantify how much of the model's output space has been explored in ways that are relevant to safety—either by testing the \emph{severity} of constraint violations (\ac{VO-KMVP}) or the \emph{diversity} of stimulation outputs (\ac{VO-KMOC}).

\ac{VO-KMVP} quantifies how thoroughly the tests explore the range of each safety constraint. 
For a given output vector $\mathbf{y}$, each safety constraint $V_k$ can be expressed as an inequality 
$V_k(\mathbf{y}) = \alpha(\mathbf{y}) - c \le 0$, 
where $\alpha(\mathbf{y})$ is a biophysical quantity of interest (for example, charge density or total current) 
and $c$ is the physiological limit of that quantity. 
The ratio $\alpha(\mathbf{y})/c$ is therefore a dimensionless \emph{violation proportion}: 
values below~1 correspond to safe stimulation, while values at or above~1 indicate a violation.

Two types of constraints are considered (see Section~\ref{sec:violations}): 
aggregate constraints $V_A$ that depend on all electrodes jointly and electrode-wise constraints $V_E$ that apply separately to each electrode $i \in \mathcal{I}$. 
For each type, the violation proportions are divided into $K$ equal-width bins over a range $[\textit{min}, \textit{max}]$, 
and a bin is considered ``covered'' once at least one test has produced a value in that bin's range.
The coverage of a test set $S$ is then defined as

\begin{equation}
\textsc{Cov}(S) = 
\frac{\textsc{Part}(V_A) + \textsc{Part}(V_E)}
{K \times |V_A| + K \times |V_E| \times |\mathcal{I}|},
\end{equation}

where
\[
\textsc{Part}(V_A) = 
\sum_{v \in V_A}\sum_{k=0}^{K-1}
\left[\exists \mathbf{x} \in S : 
\textit{min}_{k} \leq \frac{\alpha_v(\mathbf{y})}{c} < \textit{min}_{k+1}\right],
\]
\[
\textsc{Part}(V_E) = 
\sum_{v \in V_E}\sum_{i = 1}^{|\mathcal{I}|}\sum_{k=0}^{K-1}
\left[\exists \mathbf{x} \in S :
\textit{min}_{k} \leq \frac{\alpha_{i,v}(\mathbf{y})}{c} < \textit{min}_{k+1}\right].
\]

Here $[\cdot]$ denotes the Iverson bracket (equal to~1 if the condition inside is true and 0 otherwise), 
and $\textit{min}_k = \textit{min} + k (\textit{max}-\textit{min})/K$ defines the lower edge of bin $k$. 
Values below the minimum or above the maximum are assigned to the outermost bins.
Intuitively, this metric rewards new tests that drive stimulation parameters closer to the safety boundary, helping the fuzzer find the most severe violations.

\ac{VO-KMOC} complements \ac{VO-KMVP} by measuring how broadly the tests explore the model's output space, 
irrespective of whether they cause violations. 
Let $\mathcal{O}$ denote the set of output dimensions (e.g., all electrode amplitudes, frequencies, and pulse widths). 
For each output dimension $o \in \mathcal{O}$, 
the observed range between the minimum $\textit{lo}_o$ and maximum $\textit{hi}_o$ values across a profiling dataset 
is divided into $K$ bins with boundaries 
$\textit{lo}_{o,k} = \textit{lo}_o + k (\textit{hi}_o - \textit{lo}_o)/K$.
A bin is marked as covered if at least one test input $\mathbf{x}$ produces an output $\textit{val}(\mathbf{x}, o)$ within that interval:

\begin{equation}
\textsc{Cov}(S) = 
\frac{\sum_{o \in \mathcal{O}}\sum_{k=0}^{K-1}
[\exists \mathbf{x} \in S : \textit{lo}_{o,k} \le \textit{val}(\mathbf{x}, o) < \textit{lo}_{o,k+1}]}
{K \times |\mathcal{O}|}.
\end{equation}

This metric rewards exploration of new amplitude, frequency, or pulse-duration ranges and helps identify diverse but safe operating points. 
Together, \ac{VO-KMVP} and \ac{VO-KMOC} balance \emph{depth} (i.e., how far into dangerous territory the model can go) and \emph{breadth} (i.e., how widely it explores the possible stimulation space). 
Because both metrics operate directly on physical output values, their results can be interpreted in clinically meaningful units (e.g., microamperes or microcoulombs/cm\textsuperscript{2}).

\paragraph{Additional Violation-Focused Metrics}
The remaining novel metrics (\ac{VO-KMVP}-V, VO-VCC, I-KMIC, and I-Div-Approx) extend this logic to alternative forms of diversity or constraint specificity. 
Their conceptual roles are listed in Table~\ref{tab:coverage_metrics}, 
and their formal definitions are provided in Appendix~\ref{sec:app-vo-metrics}. 
All coverage scores are computed between~0 and~1, 
with higher values indicating more complete exploration of the model’s input–output–violation space.

\subsubsection{Neuron-Coverage Metrics (White-Box Baselines)}
\label{sec:metrics-neuron}

White-box coverage metrics quantify how thoroughly a test set activates the \emph{internal neurons} of a model. 
They are termed \emph{white-box} because they require direct access to the model's internal activations, analogous to inspecting which lines of code were executed during a software test~\citep{pei2017deepxplore,ma2018deepgauge}. 
In contrast, our proposed violation-focused metrics operate in a \emph{black-box} setting, relying only on model inputs and outputs.

We implemented five representative white-box metrics from prior work in deep neural network testing and verification~\citep{pei2017deepxplore,ma2018deepgauge,li2019structural,yang2022revisiting,dong2020empirical}. 
These metrics remain the most widely used baselines in the literature and therefore provide a meaningful comparison for our black-box, safety-driven approach. 
Their names and conceptual roles are summarized in Table~\ref{tab:coverage_metrics}, and their mathematical definitions are given in Appendix~\ref{sec:app-vo-metrics} for completeness.

In essence, these metrics assess how much of a model's internal computation has been exercised by the current test set:
\begin{itemize}[topsep=0pt,itemsep=-1ex,partopsep=0pt,parsep=1ex]
    \item \emph{Neuron Coverage (NC)}~\citep{pei2017deepxplore}: counts how many neurons become active above a fixed threshold at least once.
    \item \emph{K-Multisection Neuron Coverage (KMNC)}~\citep{ma2018deepgauge}: divides each neuron's activation range into $K$ bins and measures which bins are hit, promoting exploration of intermediate activations.
    \item \emph{Neuron Boundary Coverage (NBC)} and \emph{Strong Neuron Activation Coverage (SNAC)}~\citep{ma2018deepgauge}: reward tests that drive neurons below or above their activation limits observed during training.
    \item \emph{Top-K Neuron Coverage (TKNC)}~\citep{ma2018deepgauge}: measures how often each neuron ranks among the $K$ most active units within its layer.
\end{itemize}

Together, these methods represent the current state of white-box testing in machine learning and remain important historical baselines. 
They test whether increasing the diversity of internal activations correlates with externally observable safety violations. 
However, as shown in prior analyses~\citep{yang2022revisiting,li2019structural,dong2020empirical,huang2024neuron}, neuron coverage has limited predictive value for real-world robustness, highlighting the need for output-level, biophysically grounded metrics like those introduced here.

\subsection{Measurement of Violation Diversity}
\label{sec:diversityformal}

Simply tallying discovered violations gives a false sense of progress. 
Consider a model that outputs a pulse width larger than the inverse of its frequency so the pulse cannot physically exist.
A mutation strategy can be to find one such input and then generate thousands of tiny variations that all trigger the same impossible pulse. 
The result is a huge violation count, but no insight into the extent of violations across model behaviors. 
To address this, we measure not only \emph{how many} violations are found but \emph{how} those violations are \emph{distributed} in the input feature space and across the electrode array, so we can tell whether failures are concentrated, trivial to reproduce, or genuinely diverse and actionable.

\subsubsection{Input-Feature Diversity (Geometric Diversity).}
\citet{aghababaeyan2023black} proposed several image-feature-based diversity metrics for analyzing classifier models and showed that they correlate more strongly with test quality and misclassification discovery than neuron coverage alone. 
Among these, \emph{Geometric Diversity (GD)} was found to perform best. 
It measures how spread out a set of images is in a deep feature space extracted from a pretrained convolutional network.

Following this approach, we compute feature vectors for each violating input image using the VGG16 model~\citep{simonyan2014very}. 
Let $\mathcal{F}$ denote the set of all extracted feature vectors, and let $A_\mathcal{F}$ represent the matrix formed by stacking these vectors. 
The geometric diversity is defined as the determinant of the Gram matrix $A_\mathcal{F} A_\mathcal{F}^T$:

\begin{equation}
\textit{GD} = \det(A_\mathcal{F} A_\mathcal{F}^T).
\end{equation}

A higher determinant indicates that the feature vectors are more linearly independent, implying that the violating inputs occupy a broader region of feature space.

\subsubsection{Violation-Space Diversity.}
While geometric diversity captures diversity in the visual input space, it does not account for how violations are distributed across the stimulation electrodes. 
For prosthetic devices, it is often more informative to ask whether violations are localized to specific electrodes or distributed across the array. 
We therefore introduce a complementary metric, \textit{Violation-Space Diversity (VD)}, which measures how violations vary across electrodes and violation types.

For each electrode $i \in \mathcal{I}$ and each electrode-specific constraint (here $V_{PI}$ and $V_{CD}$), we define a \emph{degree of violation intensity}:

\begin{equation}
   \textit{degree}_i =  
   \begin{cases} 
      0 & \alpha_i(\mathbf{y})/c - 1 \le 0, \\
      \alpha_i(\mathbf{y})/c - 1 & \text{otherwise.}
   \end{cases}
\end{equation}

This quantity equals zero for safe electrodes and increases with the severity of the violation. 
Because the remaining constraints ($V_{IC}$ and $V_{AE}$) operate globally rather than per electrode, they are not included directly. 
However, both relate to electrode amplitudes, so we also include the raw amplitudes $a_i$ in the analysis.

For each test, we construct a concatenated vector of normalized values 
$\mathbf{v}_i = [\textit{degree}_{PI,i}, \textit{degree}_{CD,i}, a_i]$ 
across all electrodes $i$, resulting in a violation-space matrix $A_{\textit{VS}}$ of size $|\mathcal{I}| \times 3$. 
We then measure the overall spread of these vectors across tests using the standard deviation (STD) procedure described by \citet{aghababaeyan2023black}:

\begin{equation}
\textit{STD} = \left\|\sqrt{\sum_{i=1}^{n}\frac{A_{\textit{VS},j} - \mu_j}{n}}\, , \ 1 \le j < |\mathcal{I}| \times 3\right\|.
\end{equation}

Here, $\mu_j$ is the mean of feature $j$ across the test set and $n$ the number of violating samples. 
A high violation-space diversity score indicates that violations occur with varying intensity across different electrodes rather than clustering in a single region, suggesting a more thorough exploration of the device's safety boundaries.

Finally, because both diversity metrics are impacted by sample size, we compute each on five randomly selected subsets of 200 violating input/output pairs and report the mean across subsets.

\subsection{Experimental Setup}
\label{sec:setup}

All experiments were implemented in Python using both PyTorch and TensorFlow frameworks with mixed-precision inference for efficiency. 
Fuzzing and coverage computations were run on high-performance NVIDIA GPUs: an A6000 for retinal models and an RTX 4090 for cortical models. 

We evaluate each coverage strategy for an equal wall-clock runtime to ensure fair comparison despite differences in computational overhead. We determine the total number of evaluated inputs by scaling the ratio between the time to execute a single test, and compute its coverage contribution relative to the time to execute a single test in our baseline strategies. 

Hyperparameters for previously published neuron-coverage metrics were adopted from prior work, while those for our new metrics were tuned on short pilot runs.
All model sizes, hyperparameters, number of tests per coverage metric, and code artifacts are available in the accompanying repository: \url{https://github.com/mara-downing/safety_violation_fuzzing_visual_prostheses}.

\subsection{Models Under Test}
\label{sec:models}

We applied our framework to two classes of state-of-the-art stimulus encoders representative of current approaches in visual prosthetics: retinal and cortical encoders. 
Both encoders perform an \emph{inverse mapping} from camera images to electrode stimulation parameters, but differ in anatomical target, output dimensionality, and training objectives.

\subsubsection{Retinal Stimulus Encoders.}
The retinal encoder evaluated in this work is a \ac{DSE} that maps a grayscale image to three stimulation parameters per electrode (amplitude, frequency, and pulse duration) across a $15 \times 15$ epiretinal array with \SI{400}{\micro\meter} spacing \citep{granley_human---loop_2023}.
Its objective is to generate spatially organized pulse patterns that approximate natural scene structure when viewed through a prosthetic.
Training followed a hybrid optimization procedure that combined image reconstruction losses with human-in-the-loop feedback, without incorporating explicit stimulation safety constraints.

To assess how different safety-oriented design choices influence encoder behavior, we also include a family of DSE variants introduced by \citet{schoinas2025evaluating}:
R-DSE-L (baseline), R-DSE-L\textsubscript{PR} (pulse-duration regularization), R-DSE-L\textsubscript{FR} (frequency regularization), R-DSE-L\textsubscript{PR+FR} (joint regularization), and R-DSE-L\textsubscript{FC} (hard frequency clipping).
For all retinal models, safety thresholds were set to $\epsilon_1 =  \SI{.628}{\micro\coulomb}$ for per-electrode charge limit, $\epsilon_2 = \SI{6}{\milli\ampere}$ for instantaneous current, and $\epsilon_3 = 100$ active electrodes.

A differentiable phosphene model provides the forward link between electrical stimuli and predicted percepts.
The version used here~\citep{granley_human---loop_2023} extends earlier models of epiretinal activation \citep{beyeler2019model, granley_computational_2021} by representing each electrode’s percept as a multivariate Gaussian whose size, eccentricity, and orientation depend on both local axon-fiber geometry and stimulation parameters. This  formulation incorporates well-established dependencies of phosphene brightness, size, and elongation on amplitude and frequency \citep{horsager_predicting_2009, greenwald_brightness_2009, nanduri_frequency_2012, beyeler2019model}.
Because Gaussian covariances are tilted along the underlying axon trajectories, the model captures anisotropic current spread and reproduces characteristic crescent-shaped percepts observed in human users \citep{hou_axonal_2024}.

Phosphene parameters are personalized using psychophysical fits for each simulated user, enabling realistic variation across the array.
Although percepts from individual electrodes are summed linearly, consistent with paired-electrode experiments showing near-independent activation of spatially separated axon bundles \citep{hou_axonal_2024}, the axon-map structure introduces nonlinear spatial interactions that depend on local fiber geometry.
The resulting combination of anatomical coupling and nonlinear stimulus dependence allows the model to approximate how encoder outputs transform into perceptual brightness and shape under realistic retinal constraints, making it a suitable testbed for evaluating safety violations in image-to-stimulation pipelines.

\subsubsection{Cortical Stimulus Encoders.}
The cortical encoder evaluated in this study is the C-Viseon model introduced in \citet{van2024towards}, which maps a target image to stimulation amplitudes on a subset of 60 electrodes from a 96-channel Utah array implanted in primary visual cortex \citep{normann_toward_2009, fernandez_visual_2021}.
In contrast to the retinal encoder, frequency and pulse duration are held constant across electrodes, which simplifies the output space but increases the importance of aggregate current constraints such as $V_{AE}$ and $V_{IC}$.
The encoder is trained end-to-end to minimize reconstruction error between predicted phosphenes and the target image, without explicit safety-oriented regularization.

To evaluate how architectural and loss-design choices influence safety, we include two variants from \citet{van2024towards}:
C-Viseon (baseline with no inter-electrode interaction),
C-Viseon\textsubscript{CL} (adds a co-localization loss that penalizes activation of adjacent electrodes),
and C-Viseon\textsubscript{COA} (incorporates an explicit coactivation model that increases effective current when neighboring electrodes are active).
All models assume a $10 \times 10$ Utah-array layout with \SI{0.4}{\milli\meter} electrode spacing, consistent with human V1 implants.

Safety limits were set to $\epsilon_1 = \SI{20.4}{\nano\coulomb}$, $\epsilon_2 = \SI{3.6}{\milli\ampere}$, and $\epsilon_3 = 30$ active electrodes.
These values match published human data from Utah-array stimulation studies \citep{fernandez2016cortivis, fernandez2018development, chen2020shape, fernandez_visual_2021, moure_deep_2025} and reflect conservative thresholds for safe operation in cortical implants.

\subsubsection{Evaluation Protocol.}
For each encoder, an input image $\mathbf{x}\in\mathbb{R}^d$ is mapped to a stimulation vector $\mathbf{y} = M(\mathbf{x})$.
For retinal models, $\mathbf{y}$ includes frequency, amplitude, and pulse duration at each electrode.
For cortical models, $\mathbf{y}$ consists of per-electrode amplitudes only.

All experiments report averages across three simulated users and six seed sets.
Seed sets include three images drawn from each model’s training distribution and three images from ImageNet \citep{imagenet} to evaluate generalization.
A summary of all model variants is provided in Table~\ref{tab:networkversions}, with full architectures linked in the accompanying GitHub repository.

\begin{table}[!h]
{
\caption{Overview of Models}
  \label{tab:networkversions}
  \begin{center}
  \resizebox{\linewidth}{!}{\begin{tabular}{lp{0.55\linewidth}r}
    \toprule
    Model & Specifics & Validation Loss\\
    \cmidrule(r){1-3}
    R-DSE & Retinal deep stimulus encoder~\citep{granley_human---loop_2023}. & 0.0500*\\
    R-DSE-L & R-DSE with larger training data and pulse duration modulating amplitude~\citep{granley_human---loop_2023, schoinas2025evaluating}. & 0.0625\\
    R-DSE-L$_{\mathrm{PR}}$ & R-DSE-L architecture with L2 pulse duration regularization. & 0.0703\\
    R-DSE-L$_{\mathrm{FR}}$ & R-DSE-L architecture with L2 frequency regularization.& 0.0537\\
    R-DSE-L$_{\mathrm{PR+FR}}$ & R-DSE-L architecture with L2 pulse duration and frequency regularization. & 0.0592\\
    R-DSE-L$_{\mathrm{FC}}$ & R-DSE-L architecture with frequency clipping. & 0.0628\\
    \cmidrule(r){1-3}
    C-Viseon & Cortical deep stimulus encoder~\citep{van2024towards} & 0.074
    \\
    C-Viseon$_{\mathrm{CL}}$ & C-Viseon trained to minimize activation of neighboring electrodes \citep{van2024towards}. & 0.083\\
    C-Viseon$_{\mathrm{COA}}$ & C-Viseon trained assuming each active electrode amplifies active electrodes nearby \citep{van2024towards}. & 0.088\\
    \bottomrule
\end{tabular}}
\end{center}
}
\end{table}

\section{Results}
\label{sec:results}

We evaluate the proposed coverage-guided fuzzing framework on both retinal and cortical visual prosthesis encoders introduced in Section~\ref{sec:models}.
For each model class, we compare all coverage strategies in terms of the number and diversity of safety violations uncovered.

Each fuzzing run begins with a set of seed images: across six tests of each coverage metric, three have seed sets drawn from the model's own training dataset, and three have seed sets drawn from ImageNet~\citep{imagenet}.
Seeds are iteratively mutated according to the selected coverage strategy (Section~\ref{sec:fuzzing}), producing candidate images that are passed through the encoder to generate stimulation parameters $\mathbf{y} = M(\mathbf{x})$.
These outputs are then evaluated against the safety constraints defined in Section~\ref{sec:violations}, and the process is continued until a fixed number of test cases have been evaluated.
For every coverage metric, we quantified both (i) the total number of unique safety violations discovered and (ii) the diversity of those violations across input features and electrodes (Section~\ref{sec:diversityformal}).

\begin{figure}[thp!]
\centering
\begin{subfigure}{0.47\linewidth}
  \centering
  \includegraphics[width=\linewidth]{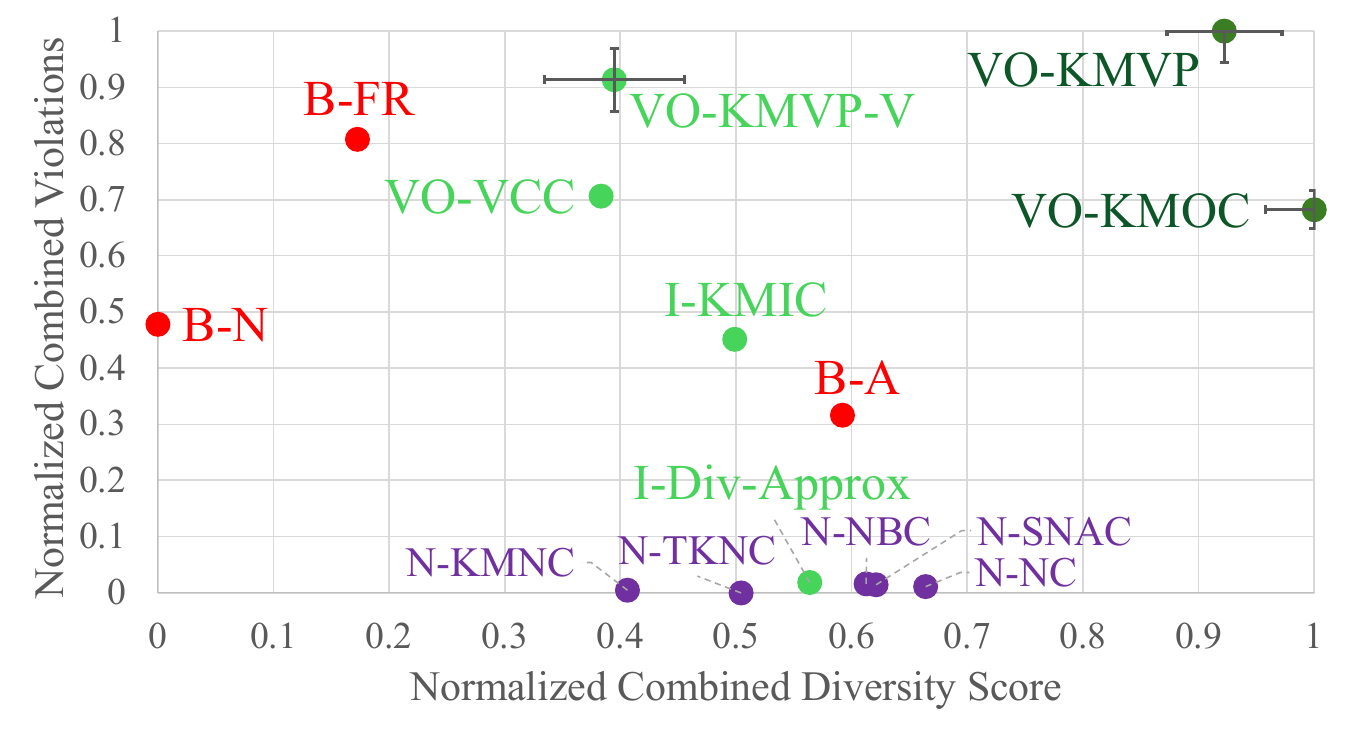}
  \label{fig:scatterplot_retinal}
\end{subfigure}%
\hfill
\begin{subfigure}{0.47\linewidth}
  \centering
  \includegraphics[width=\linewidth]{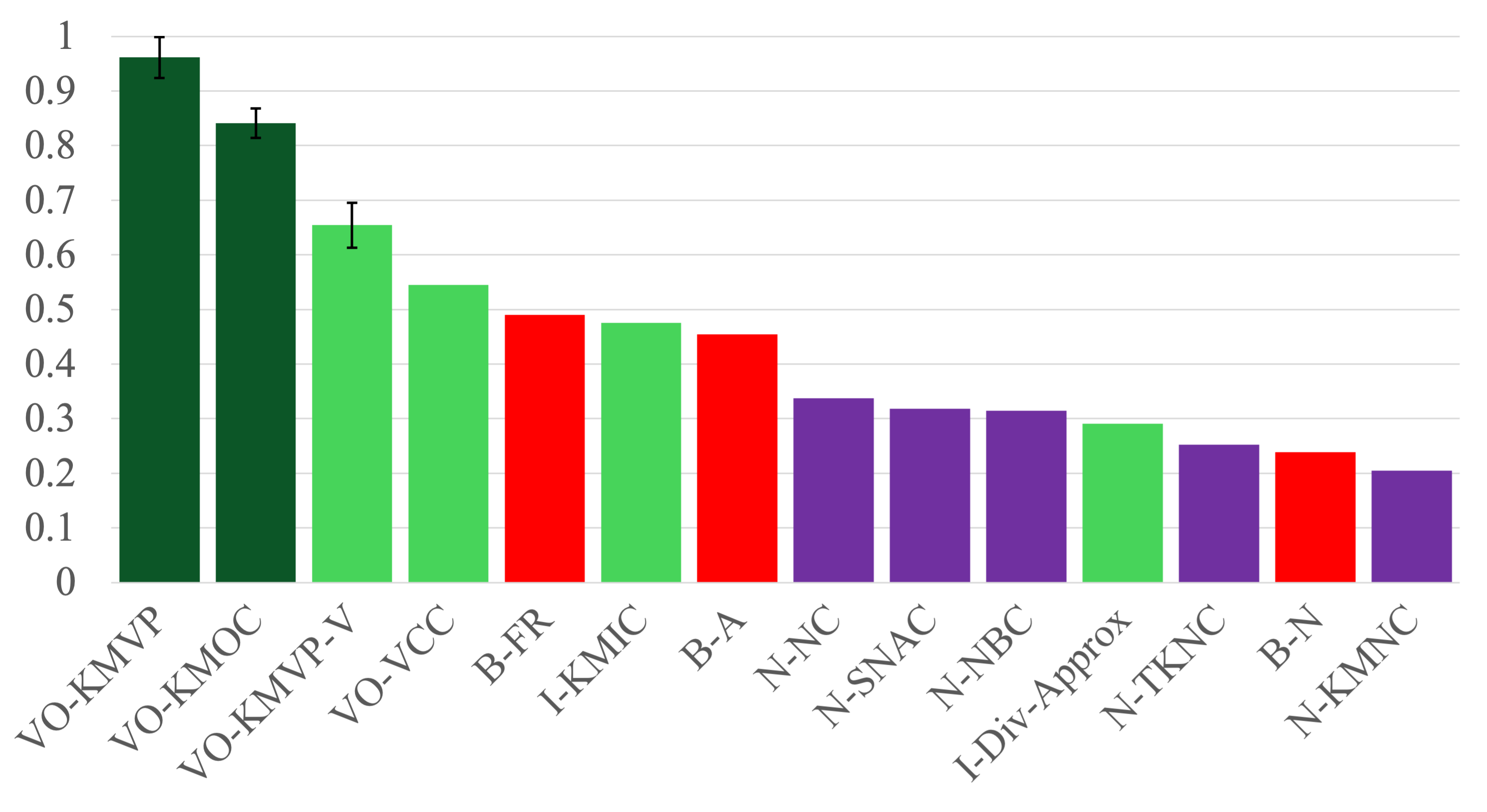}
  \label{fig:multiobjective_retinal}
\end{subfigure}
\caption{Fuzzing strategy comparison for retinal models. Left: Scatterplot of number of violations found, normalized (y-axis) and normalized combined diversity score (x-axis). Right: Normalized combination of violation and diversity score, violations and diversity equally weighted. Each datapoint is the average of six tests, each with a different seed set, and error bars show the standard error. Our metrics are shown in green, with our two best \ac{VO-KMVP} and \ac{VO-KMOC} highlighted in dark green. Neuron coverage metrics are shown in purple, and basic metrics in red.}
\label{fig:combineddivandviol_retinal}
\end{figure}

\begin{figure}[!thp]
\centering
\begin{subfigure}{.47\textwidth}
  \centering
  \includegraphics[width=\linewidth]{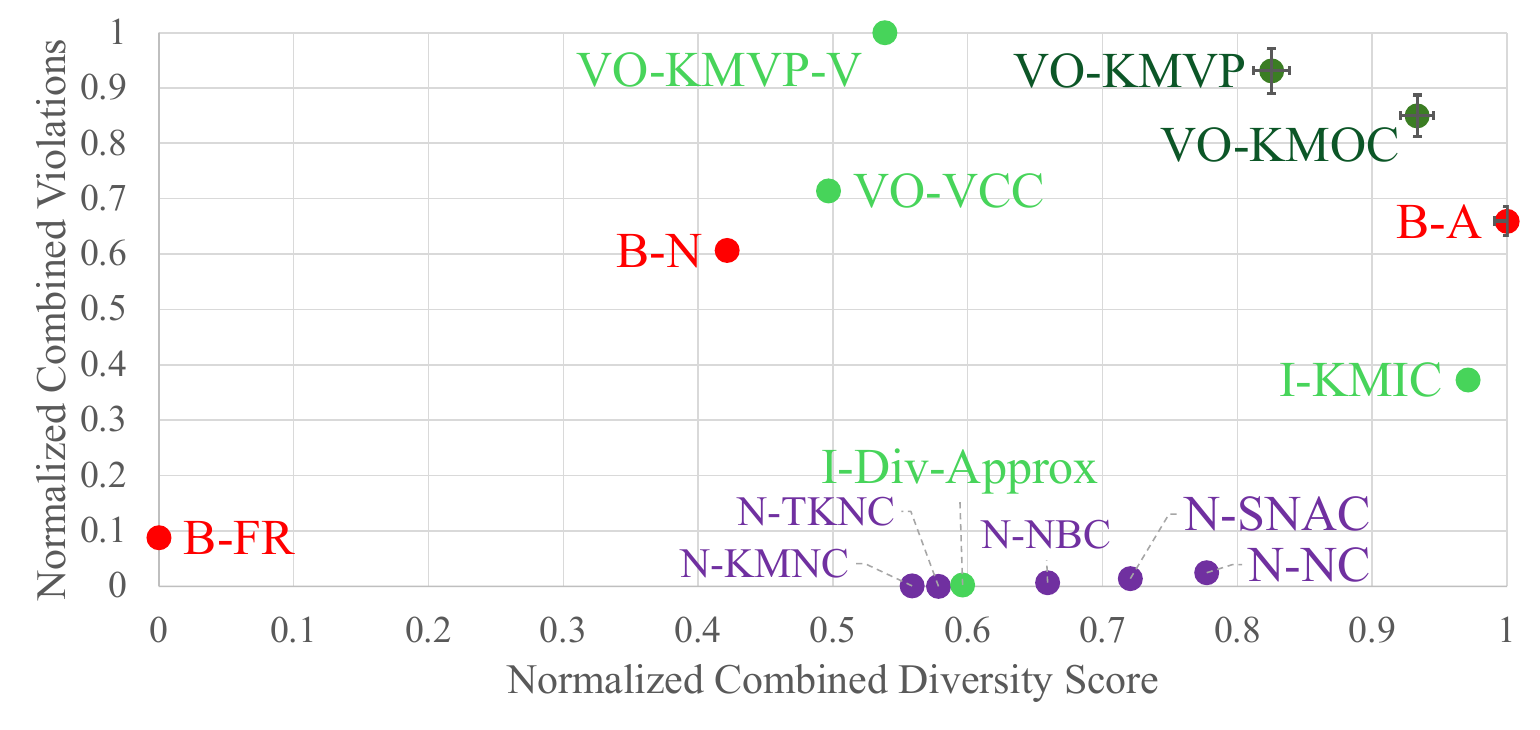}
  \label{fig:scatterplot_cortical}
\end{subfigure}%
\hfill
\begin{subfigure}{.47\textwidth}
  \centering
  \includegraphics[width=\linewidth]{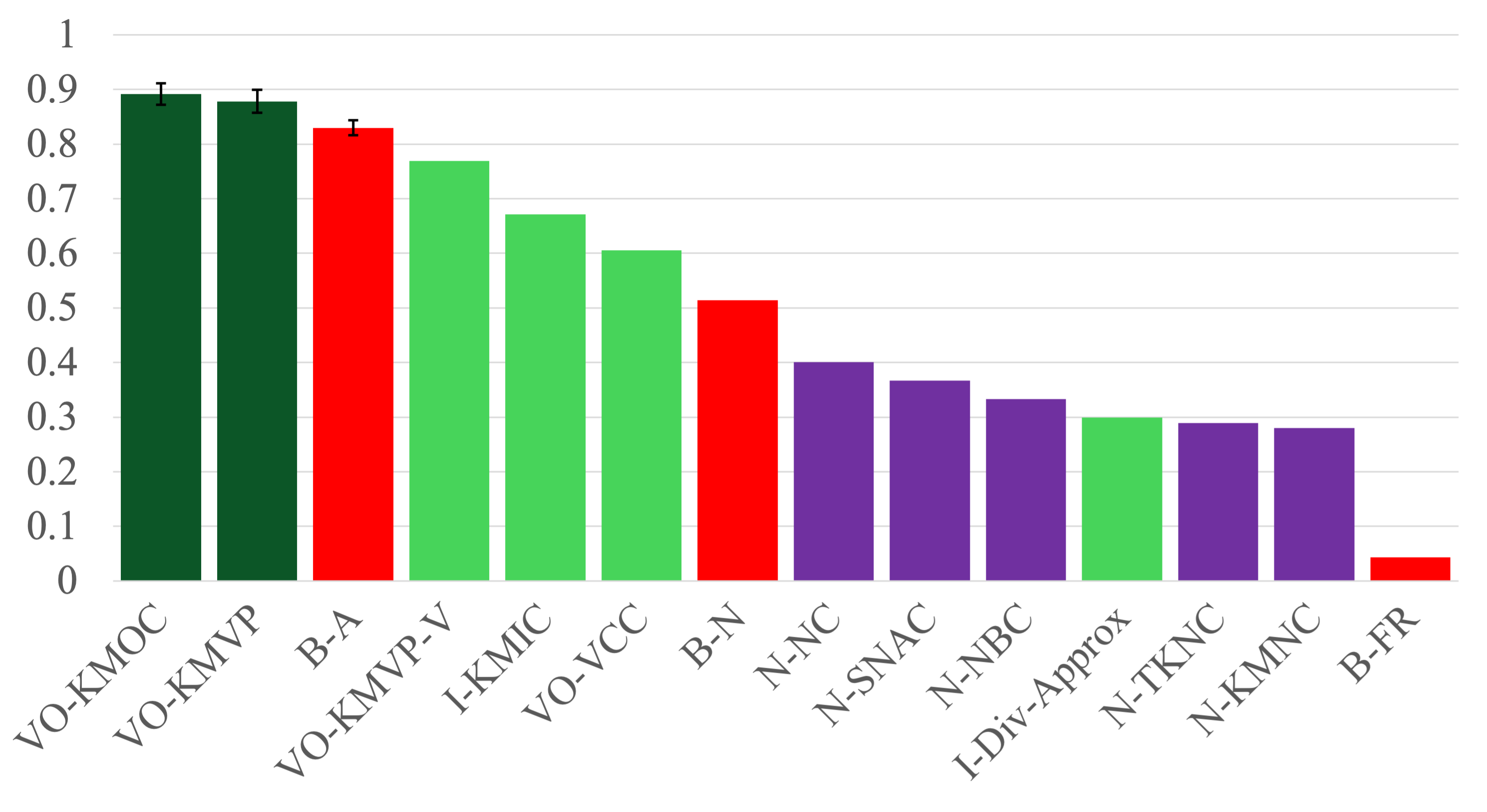}
  \label{fig:multiobjective_cortical}
\end{subfigure}
\caption{Fuzzing strategy comparison for cortical models. Left: Scatterplot of number of violations found, normalized (y-axis) and normalized combined diversity score (x-axis). Right: Normalized combination of violation and diversity score, violations and diversity equally weighted. Each datapoint is the average of six tests, each with a different seed set, and error bars show the standard error. Our metrics are shown in green, with our two best \ac{VO-KMVP} and \ac{VO-KMOC} highlighted in dark green. Neuron coverage metrics are shown in purple, and basic metrics in red.}
\label{fig:combineddivandviol_cortical}
\end{figure}

Results for the retinal encoders are shown in Figure~\ref{fig:combineddivandviol_retinal}, and results for the cortical encoders in Figure~\ref{fig:combineddivandviol_cortical}.
Each point in the left panel represents the mean across simulated participants and seed sets. Error bars show the standard error for the top three options.

For the retinal models, Figure~\ref{fig:combineddivandviol_retinal} (left) plots the number of violations discovered (y-axis) against a combined diversity score (x-axis), while Figure~\ref{fig:combineddivandviol_retinal} (right) presents the same data as a normalized composite score, equally weighting violations and diversity to highlight the best-performing metrics on a shared scale. Due to it being an outlier, we omit the B-Local basic metric, which yields a normalized combined violation score of $1.60$ and a diversity score of $-0.78$.

Results for the cortical models follow the same format (Figure~\ref{fig:combineddivandviol_cortical}), where B-Local again produces an extreme imbalance between violation count and diversity (violation score $2.80$, diversity $-0.94$).

Across both model families, violation-output coverage metrics (particularly \ac{VO-KMVP} and \ac{VO-KMOC}) discover more unique and spatially distributed violations than random or neuron-coverage baselines.
Their joint optimization of severity and output-space breadth yields higher diversity without overproducing trivial variants, indicating that coverage-guided fuzzing can expose clinically relevant failure modes in complex neurostimulation models.

\subsection{Violation Discovery for Model Selection}
\label{sec:ntwkcompare}

We next examine how violation-focused fuzzing can differentiate models trained with varying degrees of safety regularization.
Six retinal encoders and three cortical encoders were tested using the \ac{VO-KMVP} metric under identical conditions (Table~\ref{tab:networkversions}).
Each model shared the same architecture but differed in regularization terms, clipping strategies, or additional loss components designed to reduce unsafe stimulation.

\begin{figure}[thp!]
  \centering
  \includegraphics[width=0.95\linewidth]{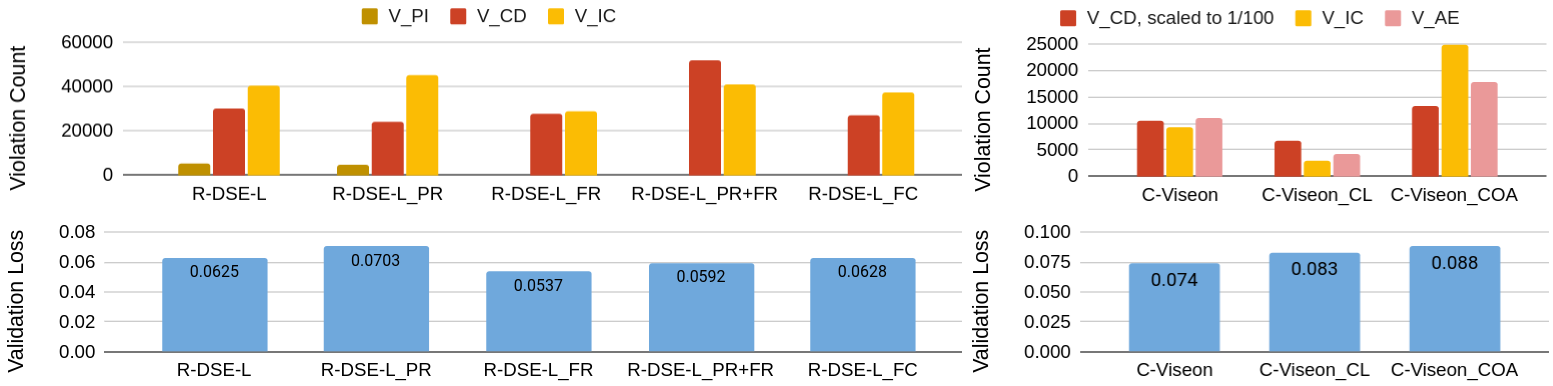}
  \caption{Bar charts of violations found and validation loss for each trained model variant (described in Table~\ref{tab:networkversions}). Left: retinal models; Right: cortical models. $V_{CD}$ and $V_{IC}$ violations are present in both retinal and cortical models, but $V_{PI}$ violations are only shown for retinal (impossible in cortical) and $V_{AE}$ violations are only shown for cortical (not discovered in retinal).}
  \label{fig:comp}
  \vspace{-10pt}
\end{figure}

Figure~\ref{fig:comp} summarizes the total number of violations discovered for each model and constraint type ($V_{PI}$, $V_{CD}$, $V_{IC}$, $V_{AE}$).
Models with explicit loss penalties on pulse duration or frequency (R-DSE-L$_\mathrm{PR}$, R-DSE-L$_\mathrm{FR}$, R-DSE-L$_\mathrm{PR+FR}$) consistently reduced the number of violations compared to the baseline R-DSE-L model, while simple frequency clipping (R-DSE-L$_\mathrm{FC}$) achieved the largest overall reduction in $V_{PI}$ violations. However, improvements were not always without penalty; for instance, regularizing pulse width and frequency (R-DSE-L$_{\mathrm{PR+FR}}$) resulted in low $V_{PI}$ violations but higher $V_{CD}$ than the R-DSE-L model, and regularizing just pulse width (R-DSE-L$_{\mathrm{PR}}$) resulted in lower $V_{CD}$ violations but higher $V_{IC}$ violations.
The unregularized R-DSE model produced catastrophic outputs, exceeding 500,000 combined $V_{PI}$ and $V_{CD}$ events, and is thus omitted from Figure~\ref{fig:comp}.

For cortical encoders, coactivation-aware and lateral-inhibition models (C-Viseon$_\mathrm{COA}$, C-Viseon$_\mathrm{CL}$) lowered the rate of overstimulation violations ($V_{AE}$) relative to the base C-Viseon model, but none fully eliminated unsafe conditions. Cortical models showed lower counts of $V_{IC}$ violations than retinal models and zero $V_{PI}$ violations, but higher counts of $V_{CD}$ violations and $V_{AE}$ violations.
Across both retinal and cortical architectures, the fuzzing framework successfully exposed violations that were not apparent from training or validation loss alone, illustrating its potential as a quantitative benchmark for model comparison.


\section{Discussion}
\label{sec:discussion}

This study introduces a coverage-guided fuzzing framework~\citep{chen2018systematic} for systematically uncovering output-level violations in machine-learning models for neural stimulation.
By formalizing domain-specific inequality constraints and defining violation-output coverage metrics, we provide a quantitative method to test whether trained models respect physiological limits under diverse and perturbed inputs.
Applied to deep stimulus encoders for retinal and cortical visual prostheses~\citep{granley_human---loop_2023,van2024towards}, the framework identified unsafe stimulation patterns that conventional loss functions and validation metrics failed to expose.

\subsection{Violation-Focused Fuzzing Reveals Hidden Failure Modes}
Across both prosthesis types, violation-output coverage metrics (\ac{VO-KMVP} and \ac{VO-KMOC}) provided the most informative search signals.
By quantifying exploration directly in the space of stimulation outputs, they enable black-box evaluation of trained encoders without requiring architectural access or inspection of internal activations.
These metrics consistently uncovered a broader and more diverse set of violations than neuron-coverage or random baselines, demonstrating that safety can be assessed as an observable property of the input-output mapping itself.

Because the violations are expressed in physical units, the results can be interpreted in terms of quantities that matter clinically, such as charge density, instantaneous current, and the number of active electrodes.
Framing safety in these domain-relevant units allows direct comparison against established biophysical limits and provides a clear link between model behavior and known failure modes in implantable systems.

Taken together, these results show that violation-guided fuzzing offers a practical and quantitative approach for identifying unsafe operating regimes in \ac{ML}-driven neurostimulation.

\subsection{Retinal and Cortical Insights}

For epiretinal encoders~\citep{granley_human---loop_2023}, fuzzing revealed that models optimized for perceptual fidelity can still generate physically impossible or unsafe pulses when exposed to out-of-distribution inputs.
Frequency regularization mitigated these violations, while duration penalties alone had limited effect, highlighting the nonlinear coupling between pulse width and frequency in charge accumulation~\citep{horsager_temporal_2011,nanduri_frequency_2012,ghaffari_effect_2020,hou_axonal_2024}.

In cortical encoders~\citep{van2024towards}, loss terms discouraging co-activation of neighboring electrodes improved safety margins by reducing total current and the number of simultaneously active sites, consistent with intracortical studies of spatial interference and excitability~\citep{fernandez_visual_2021,chen2020shape,moure_deep_2025}.
Conversely, models that explicitly incorporated current-spread or interaction terms inspired by experimental findings on crosstalk and waveform asymmetries~\citep{wilke_electric_2011,yucel_factors_2022,haji_ghaffari_improving_2021} tended to amplify unsafe amplitudes.

These results show that even biologically motivated modeling choices can introduce new risks if not empirically verified, underscoring the need for systematic post-training validation rather than reliance on training objectives or validation loss alone.

\subsection{Toward Principled Model Validation}

Violation-guided fuzzing reframes safety evaluation as an empirical and reproducible testing process rather than an indirect training objective.
It enables quantitative comparison of architectures, regularization schemes, and constraint formulations under identical conditions, offering interpretable robustness measures that complement perceptual or reconstruction metrics.
This perspective is particularly relevant for implantable neurotechnologies, where reliability depends not only on performance but also on demonstrable adherence to physiological limits~\citep{shannon_model_1992,merrill_electrical_2005}.
At the same time, long-term usability and trust strongly influence whether implant recipients rely on their devices in everyday life~\citep{nadolskis_aligning_2024}.
Integrating physiological validation with real-world user experience is therefore critical to translating engineering progress into functional benefit.

Prior work that invokes ``safety'' in ML-driven neurostimulation has treated it synonymously with minimizing delivered current or charge during optimization~\citep{shah_computational_2020,willis_optimizing_2025,kuccukouglu2025end}.
While such regularization can reduce mean output power, it does not verify that trained models remain within physiological limits once deployed, particularly under novel or perturbed inputs.
Our findings show that aggregate or spatially localized violations can still arise from complex pulse interactions even when average current is minimized.
By empirically testing deployed encoders and quantifying both the frequency and severity of violations, our framework complements these loss-based efforts and establishes a foundation for evidence-based safety benchmarking, which is a necessary step toward regulatory approval and ethical deployment of adaptive neural interfaces.
Beyond compliance, ensuring demonstrable safety and reliability is an ethical imperative for responsible neurotechnology and AI governance~\citep{yuste_four_2017}.

\subsection{Limitations and Future Directions}

The present study establishes a foundation for systematic, output-level safety evaluation, yet several natural extensions remain.

Our experiments rely on simulation-based encoders and safety thresholds derived from published device specifications, which provides a controlled and reproducible testbed.
The next step is hardware-in-the-loop validation, where electrode impedance, charge-transfer efficiency, thermal load, and long-term tissue response can be incorporated directly into the testing pipeline~\citep{fernandez2018development}.
Integrating these physiological factors will enable more comprehensive assessments, and using violation feedback as a differentiable signal during training may allow joint optimization of perceptual fidelity and safety compliance.

Beyond the biophysics of stimulation, functional outcomes in prosthetic vision depend on cortical plasticity and perceptual learning~\citep{beyeler_learning_2017,lunghi_visual_2019,caravaca-rodriguez_implications_2022,esquenazi_perceptual_2025}.
Extending violation analysis to include these neural and behavioral constraints could yield a more holistic framework that links device safety to perceptual outcomes and long-term usability.

Although this study focused on visual prostheses, the same methodology applies to other neuromodulatory systems which might use model-generated stimulation parameters such as deep-brain, spinal, and vagus-nerve stimulators~\citep{little_adaptive_2013,rao_towards_2019, drakopoulos2023neural, okorokova2018biomimetic}.
As neuroprosthetic systems become increasingly autonomous and incorporate co-adaptive or reinforcement-learning components~\citep{little_adaptive_2013,shenoy_combining_2014,shanechi_rapid_2017,rao_towards_2019}, output-level safety validation will be essential for closed-loop operation~\citep{grani_toward_2022,moure_deep_2025,beyeler_bionic_2025}.

By grounding evaluation in measurable, domain-specific constraints, violation-focused fuzzing provides a bridge between algorithmic innovation and clinical reliability.
As intelligent neurotechnologies advance toward adaptive and closed-loop operation, such frameworks will be vital not only for improving device performance but also for ensuring ethical, trustworthy, and regulatory confidence in next-generation neural interfaces~\citep{yuste_four_2017}.

\section{Conclusions}
This work introduces violation-focused fuzzing as a systematic approach to evaluating the safety of machine-learning-based neural stimulation.
By transforming safety from a training heuristic into an empirically measurable property, this framework enables reproducible benchmarking across architectures and regularization schemes and presents two highly effective coverage metrics for this measurement.
Applied to deep stimulus encoders for the retina and visual cortex, it uncovers unsafe behaviors invisible to traditional validation metrics and established a quantitative basis for model certification.
As neuroprosthetic systems advance toward adaptive, closed-loop operation, principled safety testing is essential to ensure both functional reliability and long-term patient trust.


\funding{Partially supported by the National Science Foundation (NSF) under Awards \#2124039 and \#2008660 to TB and by the National Library of Medicine of the National Institutes of Health (NIH) under Award Number DP2-LM014268 to MB. The content is solely the responsibility of the authors and does not necessarily represent the official views of the NSF or NIH.}

\roles{
Conceptualization: MD, JG, MB, TB.
Investigation: MD, MP, TB.
Data Curation: MD, TB.
Formal Analysis: MD, TB.
Software: MD, MP, TB.
Resources: MP, JG.
Visualization: MD, MP, TB.
Writing -- Original Draft: MD, TB.
Writing -- Review and Editing: MD, MP, JG, MB, TB.
Project Administration: MB, TB.
Funding Acquisition: MB, TB.
}

\data{Code and data are available at 
\url{https://github.com/mara-downing/safety_violation_fuzzing_visual_prostheses}.}


{
\small
\bibliographystyle{unsrtnat}
\bibliography{bibfile}
}

\pagebreak
\appendix

\section{Additional Violation-Focused Coverage Metrics}
\label{sec:app-vo-metrics}

\paragraph{\ac{VO-KMVP}-V}
\textit{K-Multisection Violation Proportion (only Violation)} coverage uses the same strategy as \textit{KMVP}, with the caveat that new coverage is only valid if the new coverage is in a bin indicating a violation ($\alpha(\mathbf{y})/c \geq 1$ or $\alpha_i(\mathbf{y})/c \geq 1$).

\paragraph{VO-VCC} \textit{Violation Constraint Coverage} uses the same strategy as \textit{KMVP}, with $K=2$ and $max = 2$, which results in a coverage computation where the bins indicate presence or absence of violation but not its degree.

\paragraph{I-KMIC} \textit{K-Multisection Input Coverage} splits each pixel's value range into $K$ equal-size bins. Let $\mathcal{P}$ be the set of pixels in the input image and let $\textit{min}$ and $\textit{max}$ be the upper and lower valid pixel values. Let $\textit{min}_{k} = \textit{min} + k \times (\textit{max} - \textit{min})/K$, then

\begin{equation}
\textsc{Cov}(S) = \frac{\Sigma_{p \in \mathcal{P}}
\Sigma_{k=0}^{K-1}[\exists \mathbf{x} \in S \  . \ \textit{min}_{k} \leq \textit{val}(\mathbf{x}, p) < \textit{min}_{k+1} 
]}{K \times |\mathcal{P}|}
\end{equation}

\paragraph{I-Div-Approx}
{\em Diversity Approximation} coverage computation begins with a profiling step similar to KMNC, NBC, and SNAC, in which the model is executed on a set of input data to determine high and low values for each feature in the input-diversity feature vector. Next, each feature's range is split into $K$ equal-sized bins. 
Output values that fall outside the expected range are marked in the highest or lowest bin, depending on whether they are above or below the range. Let $\mathcal{F}$ be the set of features in the feature vector and let $lo_{f,k} = \textit{lo}_{f} + k \times (\textit{hi}_{f} - \textit{lo}_{f})/K$, then:

\begin{equation}
\textsc{Cov}(S) = \frac{\Sigma_{f \in \mathcal{F}}
\Sigma_{k=0}^{K-1}[\exists \mathbf{x} \in S \  . \ \textit{lo}_{f,k} \leq \textit{val}(\mathbf{x}, f) < \textit{lo}_{f,k+1} 
]}{K \times |\mathcal{F}|}
\end{equation}

\noindent
with the caveat that $\textit{lo}_{f,0} \leq \textit{val}(\mathbf{x}, f)$ and $\textit{lo}_{f,K-1} > \textit{val}(\mathbf{x}, f)$ always evaluate to true (values above or below the range [$\textit{lo}_{f}$, $\textit{hi}_{f}$] are counted as coverage in the lowest or highest available bin).

\paragraph{N-NC}
{\em Neuron Coverage}~\citep{pei2017deepxplore} defines an activation condition for each neuron (if the neuron's value is greater than or equal to a threshold $t$) and for each test case $x$ computes which of the neurons in $\mathcal{N}$ have been activated by running that test case:

\begin{equation}
\textsc{Cov}(S) = \frac{\Sigma_{n \in \mathcal{N}} [\exists \mathbf{x} \in S \  . \ \textit{val}(\mathbf{x}, n) \geq t]}{|\mathcal{N}|}
\end{equation}

\noindent
Recall that $[expr]$ returns 1 if \textit{expr} evaluates to true, otherwise it returns 0.

\paragraph{N-KMNC}
{\em K-Multisection Neuron Coverage}~\citep{ma2018deepgauge} partitions each neuron's value range [$\textit{lo}_{n}, \textit{hi}_{n}$] into $K$ equal-size bins. 
Let $\textit{lo}_{n,k} = \textit{lo}_{n} + k \times (\textit{hi}_{n} - \textit{lo}_{n})/K$, then:

\begin{equation}
\textsc{Cov}(S) = \frac{\Sigma_{n \in \mathcal{N}}
\Sigma_{k=0}^{K-1}[\exists \mathbf{x} \in S \  . \ \textit{lo}_{n,k} \leq \textit{val}(\mathbf{x}, n) < \textit{lo}_{n,k+1} 
]}{K \times |\mathcal{N}|}
\end{equation}


\noindent
Note that in N-KMNC coverage, neuron values outside of the range [$\textit{lo}_{n}, \textit{hi}_{n}$] are ignored.

\paragraph{N-NBC}
{\em Neuron Boundary Coverage}~\citep{ma2018deepgauge} also uses $\textit{lo}_{n}$ and $\textit{hi}_{n}$, but contrary to N-KMNC, it measures the number of neurons that take values above $\textit{hi}_{n}$ and below $\textit{lo}_{n}$: 

\begin{equation}
\textsc{Cov}(S) = \frac{\Sigma_{n \in \mathcal{N}}([ \exists \mathbf{x} \in S. \textit{val}(\mathbf{x}, n) < \textit{lo}_{n}] + [\exists \mathbf{x} \in S \ . \ \textit{val}(\mathbf{x}, n) > \textit{hi}_{n}])}{2 \times |\mathcal{N}|}
\end{equation}


\paragraph{N-SNAC}
{\em Strong Neuron Activation Coverage}~\citep{ma2018deepgauge} uses only $\textit{hi}_{n}$ and measures the number of neurons with values above $\textit{hi}_{n}$: 

\begin{equation}
\textsc{Cov}(S) = \frac{\Sigma_{n \in \mathcal{N}} [\exists \mathbf{x} \in S \ . \ \textit{val}(\mathbf{x}, n) > \textit{hi}_{n}]}{|\mathcal{N}|}
\end{equation}


\paragraph{N-TKNC}
{\em Top-K Neuron Coverage}~\citep{ma2018deepgauge} tracks which $K$ neurons have the highest values per layer when executing with a test case $\mathbf{x}$. 
Let $\textit{top}(n,\mathbf{x},K)$ denote the set of $K$ neurons in the same layer as neuron $n$ which have the $K$ top (highest) values in that layer for test input $\mathbf{x}$, then:  

\begin{equation}
\textsc{Cov}(S) = \frac{\Sigma_{n \in \mathcal{N}} [\exists \mathbf{x} \in S \  . \ n \in \textit{top}(n,\mathbf{x},K)]}{|\mathcal{N}|}
\end{equation}

\end{document}